\documentclass[article]{revtex4}
\usepackage{amssymb,amsmath,amsfonts}
\usepackage{graphicx}

\begin{document}

\title{Modeling Arcs}




\author{ Z. Insepov, J. Norem$^*$}
\affiliation{Argonne National Laboratory, Argonne, IL 60439, USA}
\author{S. Veitzer, S. Mahalingam}
\affiliation{Tech-X, Corp., Boulder CO 80803}
 \email{norem@anl.gov}




\begin{abstract}
 Although vacuum arcs were first identified over 110 years ago, they are not yet well understood.   We have since developed a model of breakdown and gradient limits that tries to explain, in a self-consistent way: arc triggering, plasma initiation, plasma evolution, surface damage and gradient limits.  We use simple PIC codes for modeling plasmas, molecular dynamics for modeling surface breakdown, and surface damage, and mesoscale surface thermodynamics and finite element electrostatic codes for to evaluate surface properties.  Since any given experiment seems to have more variables than data points, we have tried to consider a wide variety of arcing (rf structures, e beam welding, laser ablation, etc.) to help constrain the problem, and concentrate on common mechanisms.  While the mechanisms can be comparatively simple, modeling can be challenging.



\end{abstract}

\maketitle


\section{Introduction}

The field of vacuum arcs is very old \cite{earhart,lordk,alpert}.  It is complicated by the large number of variables, their large parameter range and the number of mechanisms involved, compared to the limited range of measurements that are done on individual arcs.  This work began with the goal of understanding the production of dark currents in rf cavities, where x ray production would interfere with a planned particle beam experiment \cite{PR1, moretti, insepov, hassanein, ar}.  We found, however, that by cleanly looking at the field emission from asperities at the breakdown limit, we could directly study the environment of the pre-breakdown surface.   Initial effort produced a model of the breakdown trigger based on Coulomb explosions, and subsequent work has examined the properties of arcs, primarily by assuming that unipolar arcs are the dominant mechanism \cite{Robson, schwirzke, anders1}.

\section{The Arc model}

We divide the problem of vacuum arcs into four parts: the trigger, plasma initiation,  the plasma growth phase,and  plasma damage of the metallic surface.  At all stages the driving mechanism seems to be the plasma and surface electric fields,  and Ohmic heating of the surface is not required either by the trigger or for plasma evolution.  We believe that Coulomb explosions and unipolar arcs can explain much of the behavior of these arcs.

\subsection{ Triggers}
Our work measuring dark currents and x rays from rf cavities showed that rf structures operate in a mode where the dark currents $I$ depend on the surface electric field $E$ like $I \sim E^{13} - E^{14}$ independent of state of conditioning, frequency or stored energy.  When the Fowler-Nordheim currents are plotted in this way we find that this behavior is characteristic of field  emission from a surface at a field of $\sim$10 GV/m.  At this field the electric tensile stress. $\sigma = \epsilon_0 E^2/2$ ($\epsilon_0$ being the permittivity of free space) from the electric field is comparable to the tensile strength of the material, and the material is subject to fatigue failure due to the rf oscillations.   

\begin{figure}
  \includegraphics[height=0.56\textheight]{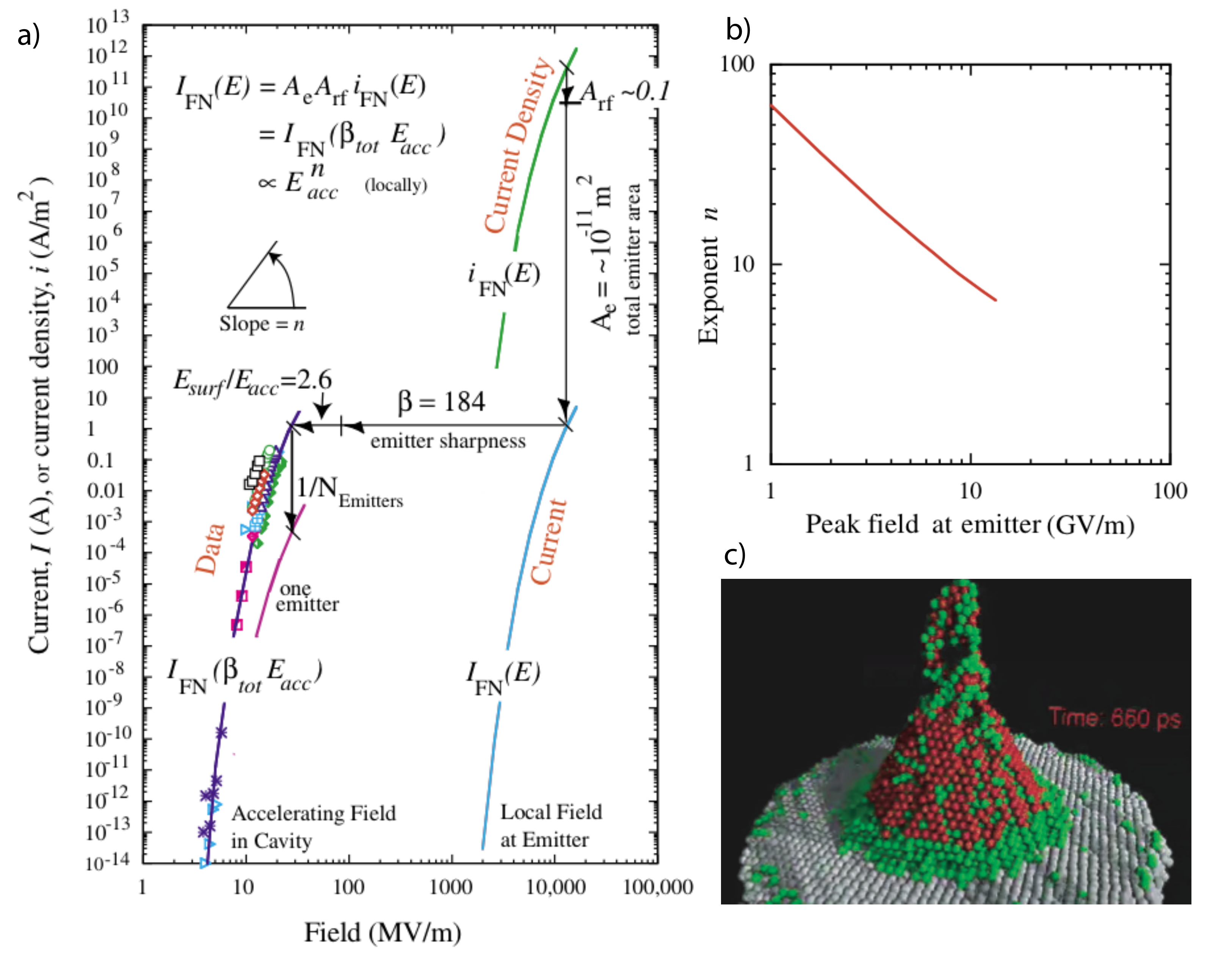}  
  \caption{a) Measurements of dark currents from a 1.3 GHz accelerator cavity showing comparisons with Fowler-Nordheim theory, with enhancement factors and emitter area factors \cite{PR1}. b) Determining the surface field from the exponent $I \sim E^n$, for a work function of 4 eV. c) Molecular Dynamics simulation of Maxwell stresses pulling copper asperities apart, which would trigger the arc \cite{insepov}.}
\end{figure}

While the conventional wisdom is that arc triggers are due to Ohmic heating in the surface, asperities of the required dimensions have not been found, and arc triggers occur randomly, with no time to warm up.  Coulomb explosions,  augmented by fatigue, do not significantly constrain the breakdown site geometry.  Note that while data shows that the breakdown rate is proportional to $\sim E^{30}$,  this behavior is compatible with a) Ohmic heating, b) failure due fatigue stress and c) electromigration, so is inconclusive.

\subsection{Plasma Initiation}
The defining property of a vacuum arc is that it can occur in a vacuum. Many experiments over the years have shown that the arc can occur at very large gaps, implying a single surface phenomenon \cite{alpert}. These conditions imply that field emitted beams must ionize material fractured off the surface by electric tensile stress.  In order to efficiently ionize material close to the cathode, our OOPIC PRO simulations show that this requires a total mass of material near the field emitter equivalent to half of a monolayer \cite{OOPIC}.  This density of gas is sufficient to accommodate a plasma density of $10^{24}$   m$^{-3}$.   

The ions produced are driven from their production point by  space charge fields, that produce a streaming, almost mono-energetic, ion current that hits the surface producing a variety of effects.

\subsection{Plasma Evolution}
Once some plasma exist above the surface the surface electric field will be affected by both the applied electric field and the surface field produced by the sheath \cite{juttner1}.  As the density, $n$, increases, simulations show that a number of things happen: 1) the sheath potential, $\phi$ stays roughly constant, since this potential is a function of the overall net charge, which evidently remains constant,  2) the surface electric field increases as $E \sim \phi / \lambda_D \sim \phi \sqrt{n}$, where $\lambda_D$ is the Debye length, 3) the increase in the surface electric field further increases the field emission current and the number of emitters.  We find that the codes show exponentially increasing arc densities, and believe that the arc eventually becomes  a non-Debye plasma.  Our model the development of rf arcs is shown in Fig 2.  We expect that DC arcs would follow a similar path in parameter space.

OOPIC Pro simulations of show that the sheath potential is on the order of 75 volts during field emission, however this is evaluated for a small, roughly spherical plasma in contact with the surface and is due to the local ionization.  This potential decreases in all directions, however, so the model should be consistent with lower burn voltages seen in small gap arcs. 

In accelerators there seem to be two limiting cases for arc interactions with their environment, which we have described as {\it killer} arcs and {\it parasitic} arcs, where the killer arcs are able to directly short out the driving potential and extinguish themselves, and parasitic arcs are unable to dispose of sufficient energy to perturb the driving field, and can survive for as long as external potentials are available.

We find that the growth time of the plasma are consistent with measurements we can make of fully formed arcs using x rays, however, the computer simulations stop before experimentally measurable currents are produced, so some extrapolation is required.

\begin{figure}
  \includegraphics[height=.3\textheight]{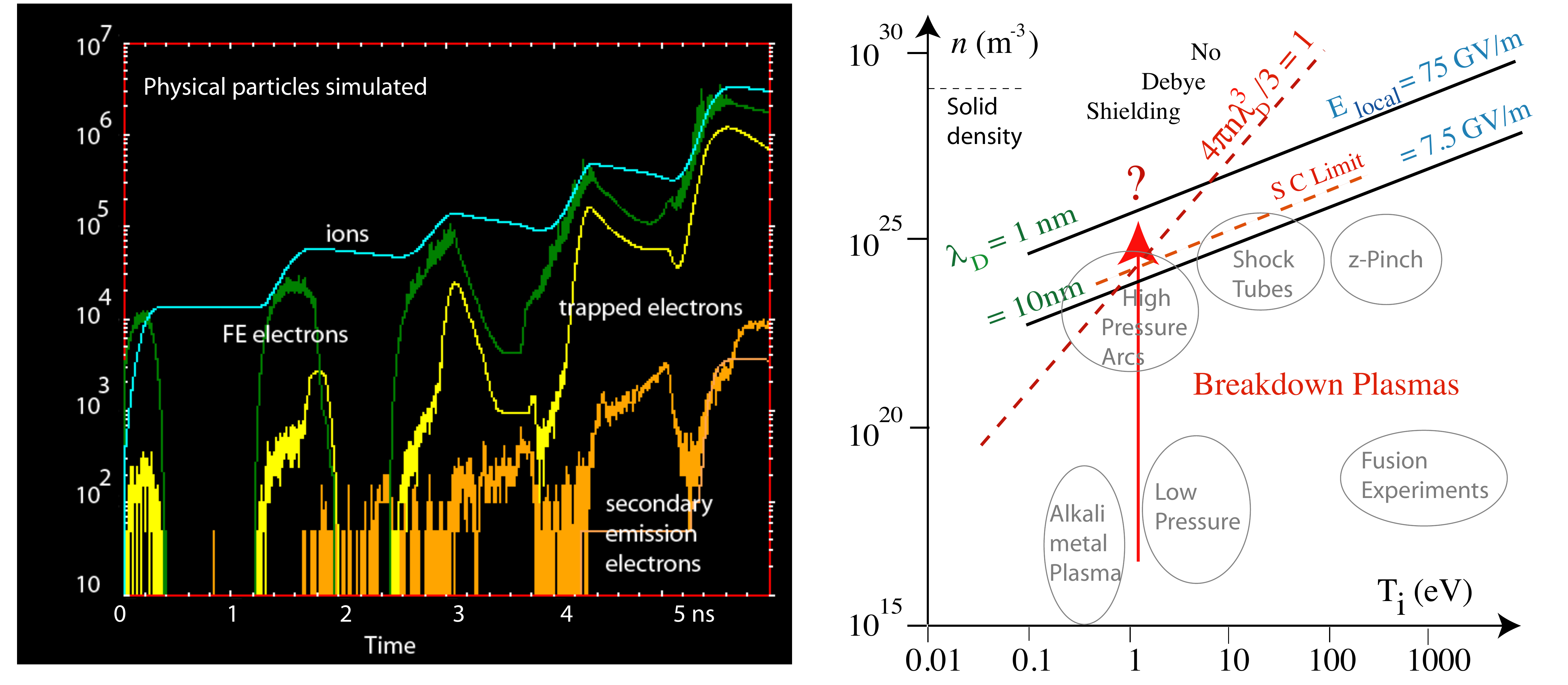}   
  \caption{Time evolution of rf breakdown at 805 MHz and the temperature (left) , and density of the plasma as calculated by OOPIC Pro (right).}
\end{figure}

\subsection{Surface Damage}

Surface damage by arcs can take many forms, depending on the parameters of the plasma and the properties of the surface.  We find that the dominant interaction is self sputtering by plasma ions streaming out of the plasma.  As the plasma evolves it becomes denser and the Debye length shortens, increasing the surface field and field emission currents and ion flux (Fig 1b.).  We have calculated the self sputtering rates for high surface fields, high surface temperatures and varying grain orientation with the results shown in Fig 2.   

Although the temperature of the plasma is very low, the potential induced by ionization produces a very high plasma pressure for the high density plasma.  After the surface is melted, the plasma pressure $p = nkT$, or $nkE$, pushes on the surface, the electrostatic field pulls on the plasma and the surface tension tends to flatten the surface.  The combination of a pulling (or pushing ) force on the surface, combined with the surface tension produces an instability in an initially flat surface which results in a spinodal  decomposition - the formation of ripples in the surface.  The dimensions of these ripples can then be used to estimate the parameters of the  plasma.  Ripples can also be produced by an oblique flow of ions carrying momentum to the surface in the same say as wind can produce ocean waves.  Thus ripples of two types can be produced on melted surface.

The high plasma pressure can produce sufficient liquid motion to generate particulates with sufficient mass and volume to uniformly cover the surface with secondary breakdown sites. 

\subsubsection{Self-sputtering}
The intense flux of low energy ions hitting the surface is the primary mechanism by which the plasma affects the surface.  In addition the deposited heat and pressure, self-sputtering helps to determine the evolution of the plasma by determining the atomic fluxes back to the plasma. Anders has shown that self-sputtering coefficients significantly above 1 (we assume 10) can produce a self sustaining arc \cite{anders1}.  Since the surface environment beneath the plasma is poorly understood, we have used Molecular Dynamics to estimate the self-sputtering coefficients, as a function of surface temperature, surface electric field and grain orientation.

\begin{itemize}
\item The effect of temperature is the simplest.  As the temperature approaches the melting point, the surface binding energy decreases and the sputtering yield increases.  Because surface atoms are more mobile slightly below the melting point, this increase occurs slightly below the bulk melting temperature as shown in Fig 3b.

\item The tensile stress induced by a strong electric field tends to pull atoms out of a surface.  In field evaporation at fields of $\sim$30 GV/m, this field is sufficient to pull individual atoms out of a polished surface.  We find, using Molecular Dynamics, that fields on the order of 1 - 3 GV/m, which we expect beneath plasma, are sufficient to significantly modify the low temperature self sputtering coefficient.

\item Surface grain orientation also seems to also affect self-sputtering yields.  Recent studies of rf systems has shown surface damage that seems to be dependent on grain orientation.  We have also calculated the sensitivity of the self-sputtering coefficient on grain orientation.  We find that this effect can be quite significant at low energies.
\end{itemize}

\begin{figure}
  \includegraphics[height=.27\textheight]{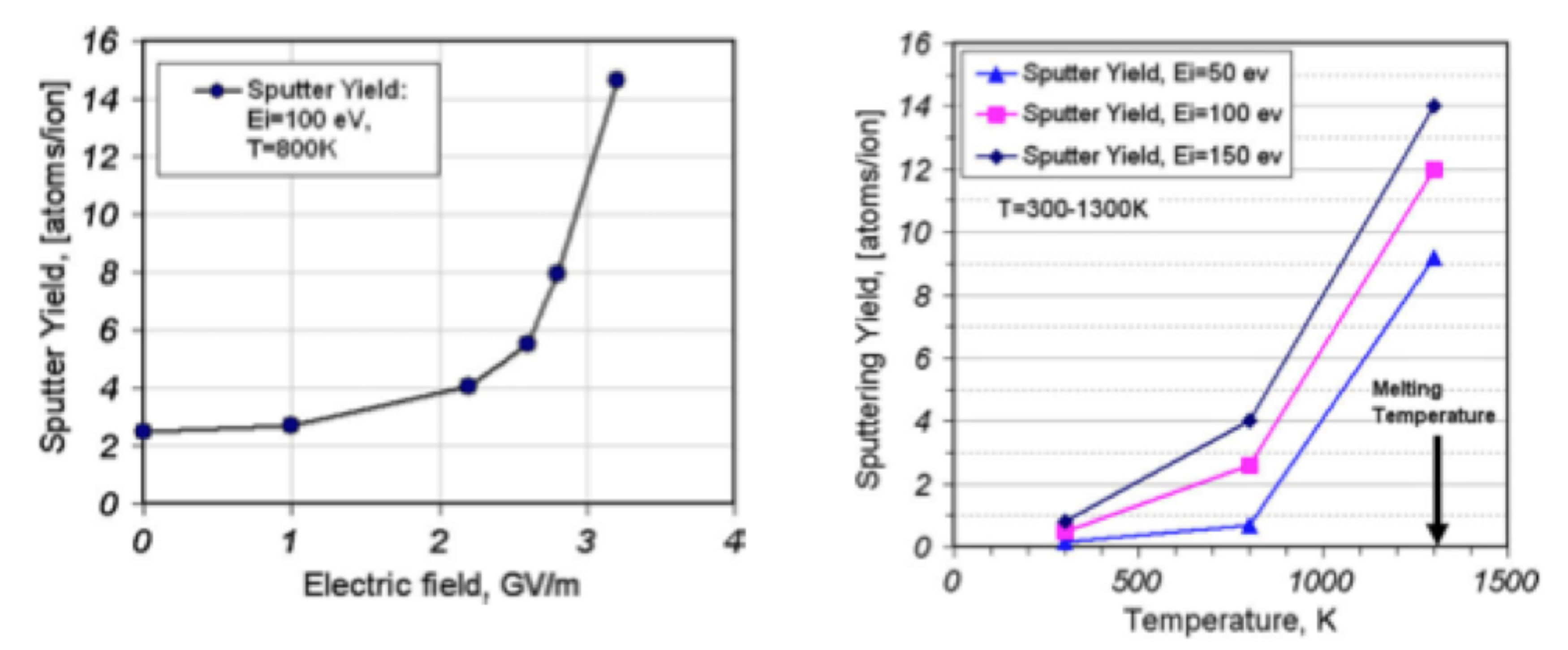}   
  \caption{Self sustained arcs require high self sputtering yields.  Calculations show either high surface fields (left) or surface temperatures (right) will provide these high yields.}
\end{figure}
 
\subsubsection{field enhancements}
Although we, like others, do not find whiskers on our surfaces, we do find many cracked regions and many sharp edged craters.  Fig 3 shows an example of one array of cracks consistent with the surface cooling by about 1000 deg.  When we model the field pattern on crack junctions, we find that the crack junctions have a field enhancement on the order of 100, which is consistent with the values we measured from dark currents.  This enhancement would be multiplied by a factor determined by the local shape of the surface, so field enhancement as large as 1000 are in principle possible. In this model the area of the emitters would be only a few nm$^2$, but a large number of them should be produced.  This analysis is described in Fig 3.   Data showing large area emitters can be accommodated by assuming that many emitters are detected.  Note that if the current is proportional to $E^{14}$, small systematic errors in measuring the electric field would produce much larger errors in measuring the emitter area, particularly if a proper two dimensional ($\beta$, area) fit of the data was made.

The conventional wisdom is that field emitters heat, vaporize and ionize due to Ohmic heating due to field emission currents.   With corners or conical geometries, the field emitting region is very small  (nm$^3$) and the thermal diffusion volume for nanosecond timescales is much larger ($\mu$m$^3$), so the heat is quickly removed and there is no significant temperature increase.  This effectively prohibits Ohmic heating as a breakdown trigger in many  cases.

\begin{figure}
  \includegraphics[height=.18\textheight]{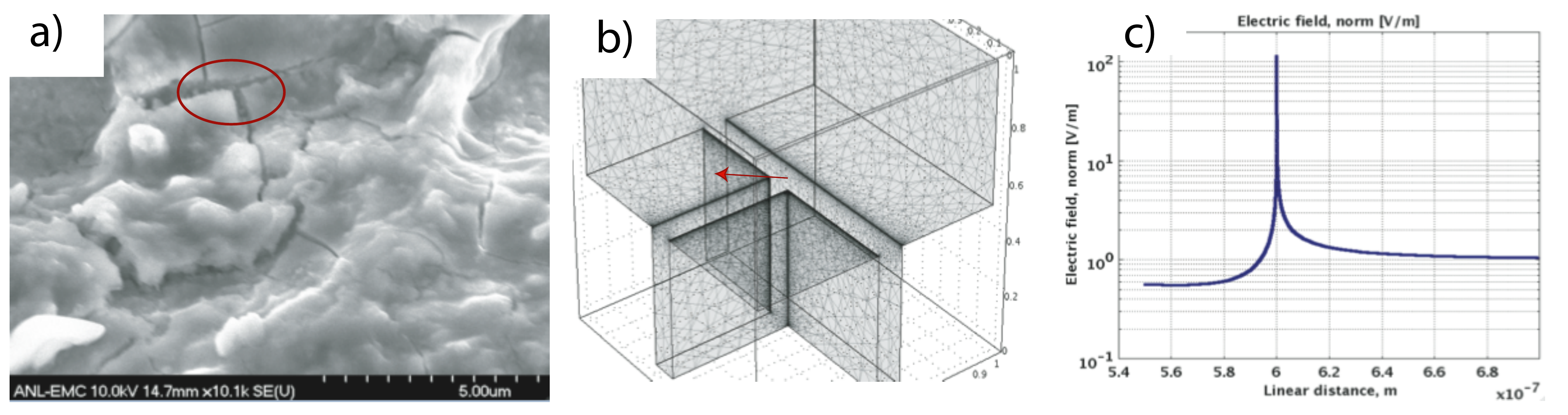}   
  \caption{a) SEM image of sub micron cracks. b)  Computer model of these crack junctions, c) Enhancement factors of electric field at the tip of these crack corners.}
\end{figure}

\begin{figure}
  \includegraphics[height=.32\textheight]{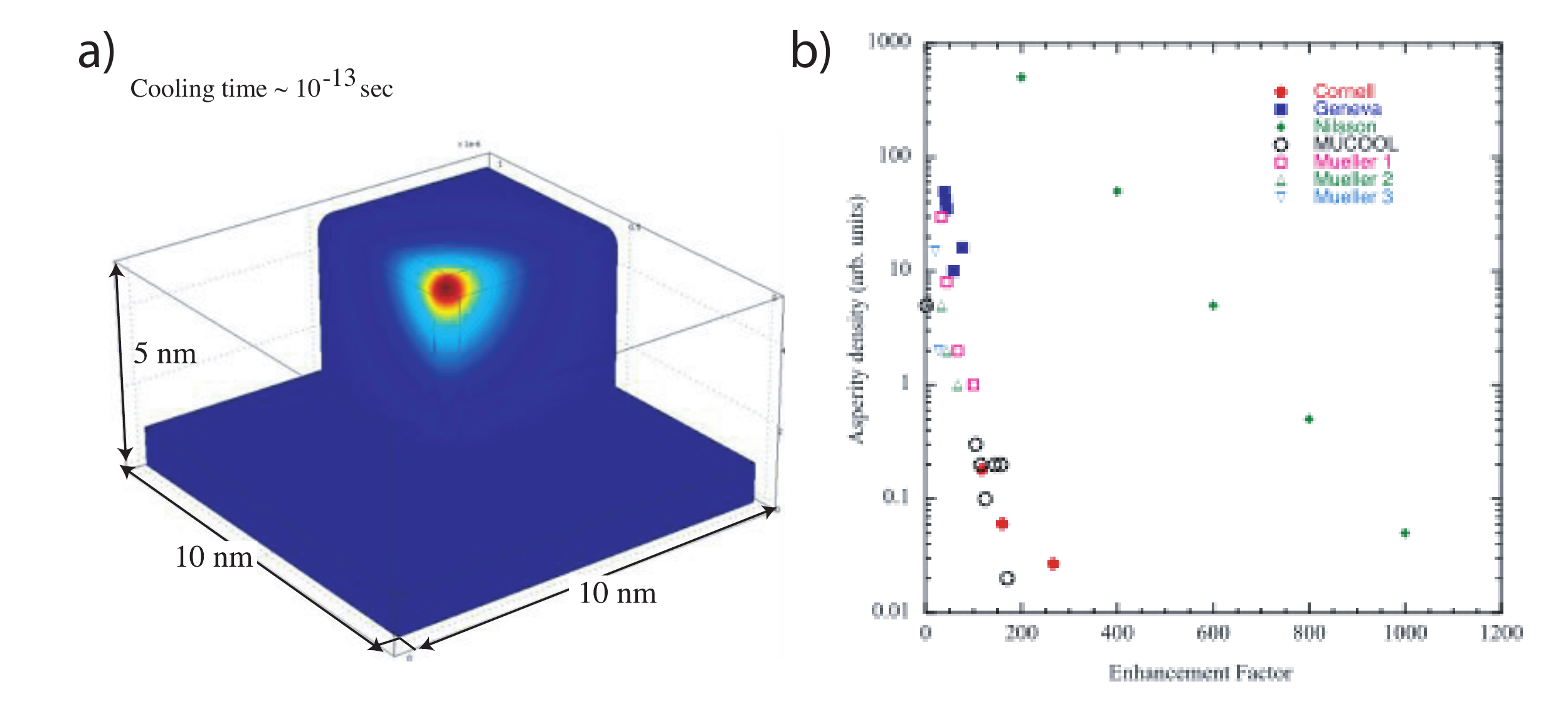}   
  \caption{Numerical simulation of heating of a blunt corner during field emission, a).  Because of the small heated volume and large thermal diffusion volume, there is little temperature fluctuation.  The spectrum of enhancement factors from a number of different sources, b) .  The spectrum is seems to show a $e^{-\beta/c}$, dependence in many data sets \cite{hassanein}.}
\end{figure}

\section{Gradient Limits}
We find that the spectrum of field enhancements for many different kinds of experiments has an exponential form with density of sites $n(\beta) \sim e^{-\beta/c}$.  If we assume that the total number and spectrum of particulates (breakdown sites) is proportional to the energy in the arc, it is possible to argue that and equilibrium enhancement factor for a given system will be produced where the equilibrium enhancement factor $\beta_{eq} \sim ln\  {U}$, where $U$ is the energy of the arc.  We have shown that this behavior is consistent with scaling laws such as the Kilpatrick limit.

Arcing phenomena are a special case of plasma surface interactions a field under active study for many reasons.   We believe that breakdown in high gradient accelerators, surface defects from e beam welding, laser ablation, tokamak rf limits, micrometeorite impacts, and some some power grid failure modes should have many common mechanisms and is should be useful to identify and study them more systematically.

\section{Comments}
The study of arcs is complicated because the number of mechanisms involved in any experiment usually exceeds the number of experimental variables.   We believe that the basic assumptions of arc models should be carefully examined as the conventional wisdom over-constrains the problem.  For example, starting with Dyke, et al., the conventional wisdom is that breakdown is initiated from ohmic heating of whiskers or asperities of comparable geometries.  The whisker geometry is unique, because the ratio of volumes in which Ohmic heating and thermal diffusion occur are comparable.  With all other geometries heating must be much less, and a different mechanism of plasma production like Coulomb explosions, more compatible with the shapes that are actually seen in SEM images, seems to be required.
 
\subsection{R\&D issues}
There are a number of general questions.  Although arcing occurs in many different fields, it is interesting to see how much of arc behavior is common to all arcing applications.  Beyond that, there are a number of more specialized questions that may become accessible.  How do non-Debye plasmas interact with materials? What are the damage mechanisms?  How do the presence of preexisting plasmas and strong external magnetic fields in three dimensions affect the triggering and evolution of an arc? How are the equilibrium field enhancements, and gradient limit affected by these variables? What is the effect of dense neutral gas on the arc? To what extent can surface treatment mitigate or modify gradient limits?   What determines the time structure of these arcs?  These questions will test modellng.
  
  \section{Conclusions}
We have developed and modeled a picture of rf arcs in accelerators, where the basic mechanisms are Coulomb explosions and unipolar arcs.  One of the problems with this field, however, is that there frequently are alternative mechanisms that can produce satisfactory comparisons to a given set of data.  We believe it is important to both increase the precision and self-consistency of the models, but it is important to extend models to the widest range of phenomena as the most useful test of their value.  

  \section{Acknowledgements}
The work at Argonne is supported by the U.S. Department of Energy Office of High Energy Physics  under Contract No. DE-AC02-06CH11357.  The work of Tech-X personnel is funded by the Department of Energy under Small Business Innovation Research Contract No. DE-FG02-07ER84833.

 \end{document}